\documentclass[aps,cyrwin,prb,twocolumn,floats,graphicx,epsfig,euscript,amsmath,amssymb]{revtex4}
\usepackage{mathrsfs}
\usepackage[dvips]{graphicx}
\usepackage{epsfig}
\DeclareMathAlphabet\EuScript{U}{eus}{m}{n} \SetMathAlphabet\EuScript{bold}{U}{eus}{b}{n}

\def\lapprox{\,\raise0.4ex\hbox{$<$}\kern-0.8em\lower0.7ex\hbox{$\sim$}\,}
\def\gapprox{\,\raise0.4ex\hbox{$>$}\kern-0.8em\lower0.7ex\hbox{$\sim$}\,}
\def\ba{\begin{array}}
\def\ea{\end{array}}

\begin{document}

\title
{Stochastization of long living spin-cyclotron excitations in a spin-unpolarised quantum Hall system }

\author{S. Dickmann and B.$\,$D. Kaysin }

\address{Institute of Solid State Physics, Russian
Academy of Sciences, Chernogolovka, 142432 Russia}


\begin{abstract}
 In the present study we address the kinetics of long-lived excitations at zero temperature in an electronic quantum Hall system with filling factor $\nu=2$. The initial coherent state of spin-cyclotron magnetoexitons with two-dimensional wave vector ${\bf q}=0$ resulting from laser pumping is stochastized over time due to emission of acoustic phonons. The elementary emission process requires participation of two magnetoexitons, so the effective rate of phonon emission is proportional to the excitation density squared, and the stochastization process occurs nonexponentially with time. The final distribution of magnetoexitons over 2D momenta, established as a result of stochastization, is compared with equilibrium distribution at finite temperature
\end{abstract}
\maketitle
In recent years there has been a clear interest in studying long-lived exciton systems in two-dimensional semiconductor structures, obviously in connection with the prospect of observing a macroscopic quantum condensate in a dense multi-exciton ensemble (similar, for instance, to the Bose-Einstein condensate). A special place here is occupied by purely electronic spin magnetoexitons formed in the conduction band of quantum Hall systems.\cite{UFN} Due to their high symmetry (in particular, their `hole', being just a vacancy at the almost occupied electron Landau level, has a negative cyclotron mass in magnitude exactly equal to the electron cyclotron mass), the number of effective relaxation channels is substantially reduced compared to common electron--valence-hole excitons. As a result, the lifetime of purely electronic cyclotron spin-flip excitons ({CSFEs}) reaches a record magnitude, $1\,$ms, in a spin-unpolarised quantum Hall system.\cite{UFN} This CSFE at filling factor $\nu=2$ (see Refs. \onlinecite{ka84,di05,ku05}) is the subject of our present theoretical study. Now, however, we will not discuss the physics of CSFE relaxation/annihilation due to transition of the system to the ground state. We address CSFE stochastization, decay of an initial coherent multi-excitonic state, where all excitations have equal 2D momenta ${\bf q}\!=\!0$, into a diffusive incoherent state provided that the total number of excitations remains constant.

We remind that generally the CSFE in a `clean' unpolarized quantum Hall system, being a collective excitation from the $S\!=\!0$ ground state, is a $S\!=\!1$ triplet characterised by 2D momentum ${\bf q}$.  The three components of the triplet differ by the values of spin quantum number $S_z=-1,0,1$, thus being energetically gapped by Zeeman energy $|g\mu_BB|$.\cite{ka84,di05,ku05} The lowest-energy excitation in GaAs quantum Hall systems corresponds to the $S_z=1$ component and conventionally represents promotion of an electron from the fully occupied zeroth Landau level upward to the first one with a simultaneous spin flip from $s_z=-1/2$ to $s_z=1/2$. The energy of this CSFE component can be calculated in terms of expansion over ratio
$r_s\!=\!E_{\rm C}/\hbar\omega_c$
of characteristic Coulomb energy $E_{\rm C}$ to cyclotron energy
$\hbar\omega_c$ (Refs. \onlinecite{ka84, di05}). {In the case of ideally two-dimensional electrons the $E_{\rm C}$
value would be equal to $e^2/\kappa l_B$ ($\kappa$ is the dielectric
constant, $l_B$ the magnetic length), however, in a modern GaAs wide-thickness quantum well $E_{\rm C}$ is considerably smaller. The CSFE energy counted off the ground-state level is determined by formula
\begin{equation}\label{total_energy}
{E}_{\bf q}=\delta^{(0)}+{\cal E}_{q},
\end{equation}
where ${\cal E}_{q}$ is the ${\bf q}$-momentum dispersion which, calculated to the first order in $r_s$, vanishes if $q\to 0$ (see review \onlinecite{UFN} and the works cited therein).  The value $\delta^{(0)}$ is the $q\!=\!0$ magnetoexciton energy,
\begin{equation}\label{delta}
\delta^{(0)}\!\equiv\!\hbar\omega_c\!\!-|g\mu_BB|\!+\!{\varepsilon}_0,
\end{equation}
that is, it includes  the cyclotron and Zeeman energies and negative Coulomb shift ${\varepsilon}_0$ that remains nonzero even if $q\!\to\!0$. [The ${\varepsilon}_0\!<\!0$ value representing the second order Coulomb correction ($\sim\!\hbar\omega_cr_s^2$)  was calculated and experimentally measured in works \onlinecite{di05}] and \onlinecite{ku05}, respectively].

The specificity of experimental excitation is that initially, as a result of laser pumping, a long-lived CSFE ensemble appears in the vicinity of point ${\bf q}=0$ of the phase $K$-space corresponding not to the minimum but to the maximum of energy ${\cal E}_{ q}$ in the range of actual values $q\!<\!1.4 \hbar/l_B$ of the 2D momenta where ${\cal E}_q$ is negative (see Fig. 1). This `zero momentum' ensemble is subsequently stochastized, so that the main mass of excitons in the $K$-space diffuses to the vicinity of the `shallow' energy minimum at $q\!\sim\!q_m\!\approx\!0.9\hbar/l_B$ values, and finally CSFEs completely relax/annihilate therefrom. The stochastization occurs without any change of the spin state, thus, certainly, it is much faster than the total CSFE-relaxation process. However, the stochastization is also associated, like relaxation, with emission of phonons and limited by the laws of conservation of energy and momentum. In particular, the stochastization is a two-excitonic process and, therefore, in this case its rate strongly depends on the CSFE concentration.

Note that, if the electronic system were completely conservative, the diffusion associated with the departure of the CSFE ensemble from the $q=0$ maximum would be simply impossible due to impossibility to provide an energy release. The conservation is violated when the interaction with the lattice is taken into account. At the same time in the `clean' (translationary invariant) system, the one-exciton process associated with the emission of a phonon is kinematically forbidden: the equality following from the energy preservation condition, ${\cal E}_q+\epsilon_{\rm ph}\!({\bf k})={\cal E}_0\!\equiv\!0$, is never fulfilled.\cite{foot1}
Here $\epsilon_{\rm ph}$ is the phonon energy, and ${\bf k}$ is the phonon wave vector with the 2D component equal to $-{\bf q}$. For example, for bulk acoustic phonons within  approximation of isotropic dispersion law, we have $\epsilon_{\rm ph}\!=\!\hbar c_s\sqrt{k_z^2\!+\!q^2}$, where it is known that $c_s\!>\!3\,10^5\,$cm$\!/$s. Fig. 1 shows  the $-\hbar c_sq$ dependence (the blue straight line) which does not intersect the ${\cal E}_q$ dispersion curve.
\vspace{-3.mm}
\begin{figure}[h]
\begin{center}
\vspace{-0.mm}
\hspace{-0.mm}
\includegraphics*[angle=0,width=.53\textwidth]{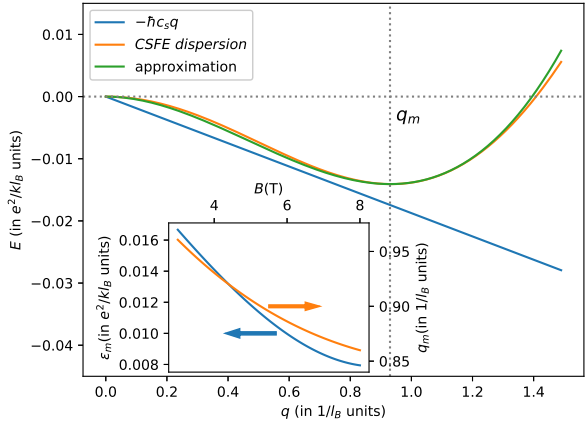}
\end{center}
\vspace{-9.mm}
\caption{In the main part: the CSFE dispersions ${\cal E}_q$, calculated at $B\!=\!4.18\,$T, in accordance with the equation \eqref{dispersion} (orange curve) and by using the analytical approximation \eqref{energy2} (green curve). The blue straight line corresponds to dependence $-\hbar c_s q$ with $c_s\!=\!3\,10^5\,$cm/s. In the inset: dependencies of ${\cal E}_m$ and $q_m$ on the magnetic field found on the base of equations \eqref{dispersion} and \eqref{formfactor}, where the $\chi(z)$ function is considered to be fixed.}
\end{figure}\vspace{-3.mm}

So, the elementary quantum transition process from the $q=0$ state to the $q>0$ one has at least to be two-excitonic. We use an approach of `excitonic representation' (for more details see Refs. \onlinecite{di05} and \onlinecite{di19}), which main idea is to abandon the basis of Fermi one-electron states and switch to the basis of so-called exciton states that diagonalize some essential part of the Coulomb interaction. The exciton states in a purely electronic quantum Hall system are generated by exciton operators originally defined via Dirac electron operators, i.e., if $p$ is the `intrinsic' quantum number of a continually degenerated Landau level and $a_p$ and $b_p$ are the annihilation operators corresponding to binary indexes $a$ and $b$ [each designates both the Landau level number and the spin sublevel, $a=(n_a,\sigma_a)$], then the exciton creation operator is$\,$ \cite{foot2}\vspace{-1.mm}
\begin{equation}\label{Q}
\displaystyle{{\cal Q}_{ab\,{\bf q}}^{\dag}={{\cal N}_{\phi}^{-1/2}}\sum_{p}\,
  e^{-iq_x\!p}\,
  b_{p+\frac{q_y}{2}}^{\dag}\,a_{p-\frac{q_y}{2}}}\vspace{-3.mm}
\end{equation}
($p$ and ${\bf q}$ are considered here and in the following to be measured in $1/l_B$ units), where ${\cal N}_{\phi}$ is the number of magnetic flux quanta in the system in question. \vspace{-1.mm}$\left(\mbox{Note also that}\, {\cal Q}_{ab\,{\bf q}}\!\equiv {\cal Q}_{ba-\!{\bf q}}^{\dag}\right)$. These ${\cal Q}$-operators have a very important property: when acting on the state of the quantum Hall system they add value $\hbar{\bf q}/l_B$ to the total momentum of the system since there occurs commutator equality \vspace{-0.mm}$\left[{\hat P},{\cal Q}_{ab{\bf q}}^{\dag}\right]={\bf q}{\cal Q}_{ab{\bf q}}^{\dag}$, where ${\hat P}$ describes the dimensionless (with $\hbar=l_B=1$) `momentum' operator (see Ref. \onlinecite{di19} and references therein). In particular, if $|0\rangle$ is the ground state, then the exciton state ${\cal Q}_{ab\,{\bf q}}^{\dag}|0\rangle$, \vspace{-.5mm} if not zero, is the eigenstate of momentum operator ${\hat P}$ with  eigen quantum number ${\bf q}$. Thus, exciton states, in contrast to single electron states, possess a natural quantum number, namely, the 2D momentum whose existence is the consequence of the translational invariance. We study CSFE at $\nu\!=\!2$, hence in our case $a\!=\!(0,\downarrow)$ and $b\!=\!(1,\uparrow)$. The ${\cal Q}_{{\bf q}}^{\dag}|0\rangle$ state (we omit the `$ab$' index, besides, note that ${\cal Q}_{{\bf q}}|0\rangle\!\equiv\!0$) represents an eigen state of our quantum Hall system yielding, in accordance with the solution of the many-electron Schr\"odinger equation,\cite{ka84,di19} the CSFE energy to the first order in terms of parameter $r_s$ with the $q$-dispersion part:\vspace{-1.mm}
\begin{equation}\label{dispersion}
{}\!\!\!{\cal E}_{q}\!\!=\!\frac{e^2}{\kappa l_B}\!\!\int_0^{\infty}\!\!\!{ds}\,e^{-s^2/2}
{\cal F}(s)\left(1\!-\!\frac{s^2}{2}\right)
  \left[1\!-\!J_0(sq)\right]\vspace{-1.mm},
\end{equation}
($J_0$ is the Bessel function). Here the formfactor is (cf. Ref. \onlinecite{an82})
\vspace{-1mm}
\begin{equation}\label{formfactor}
{}\!\!{\cal F}(q)\!\displaystyle{=\!\int\!\!\!\int\!
dz_1dz_2e^{-q|z_1-z_2|/l_B}|\chi(z_1)\chi(z_2)|^2\!,}\vspace{-1.mm}
\end{equation}
where  $\chi(z)$ describes the electron size-quantized functions in the quantum well.

We consider the situation where the CSFE ensemble represents a rarefied gas, and thus we ignore any direct interaction among excitons originating from the $e$-$e$ Coulomb coupling accounted ab initio.\cite{di19} This approach is certainly valid if the number of excitons $N$ is much smaller than ${\cal N}_\phi$ (number of electrons $\!=2{\cal N}_\phi$). However, there is an indirect inter-excitonic coupling via the GaAs lattice, which enables us to study the two-exciton process of quantum transition from the initial state\vspace{-1.mm}
\begin{equation}\label{ini}
|N\rangle=({\cal Q}^\dag_{\bf 0})^N|0\rangle\vspace{-1.mm}
\end{equation}
describing the CSFE ensemble to the final one\vspace{-1.mm}
\begin{equation}\label{fin}
|N;{\bf q}_1,{\bf q}_2\rangle=({\cal Q}^\dag_{\bf 0})^{N\!-\!2}{\cal Q}^\dag_{{\bf q}_1}{\cal Q}^\dag_{{\bf q}_2}|0\rangle\vspace{-1.mm}
\end{equation}
with nonzero momenta ${\bf q}_1$ and ${\bf q}_2$. If $-{\bf q}$ is the 2D component of the emitted phonon, i.e. ${\bf k}\!=\!(-{\bf q}, k_z)$, then\vspace{-3.5mm}
\begin{equation}\label{momenta_save}
{\bf q}_1\!+{\bf q}_2\!-{\bf q}\!=\!0.\vspace{-1.mm}
\end{equation}
In addition, the energy conservation condition must hold:\vspace{-3.mm}
\begin{equation}\label{energy_save}
{\cal E}_{q_1}+ {\cal E}_{q_1}+\epsilon_{\rm ph}({\bf k})\!=\!0.\vspace{-1.5mm}
\end{equation}

Using the excitonic representation approach, one can express any interaction, saving the number of electrons in the conduction band, in terms of ${\cal Q}$-operators of one kind or another depending on the correct choice of operators $a_p$ and $b_p$. Commutation algebra of $Q$-operators and mathematical expectations (matrix elements) containing  various excitonic and multi-excitonic states are known and have already been repeatedly calculated (see Refs. \onlinecite{di05,di19} and works cited therein). In particular, it is not difficult to find that in our case, when $1\!\ll\!N\!\ll\!{\cal N}_\phi$, the norms squared of the initial and final states are equal to
\begin{equation}\label{norms}{}\!{}\!
\begin{array}{l}
\langle N|N\rangle\approx N! \quad {\rm and}\qquad\qquad\qquad\qquad\vspace{1.mm}\\
\langle{\bf q}_2,{\bf q}_1;N|N;{\bf q}_1,{\bf q}_2\rangle\approx(N\!-\!2)!(1+\delta_{{\bf q}_1,{\bf q}_2}),
\end{array}
\end{equation}
where $\delta_{...}$ is the Kronecker delta.

Now we present the interaction of electrons with 3D acoustic phonons  in terms of the excitonic representation (see also Refs. \onlinecite{di96}, \onlinecite{di00} and \onlinecite{di12}).
The Hamiltonian is written
as (see, e.g., Ref. \onlinecite{io89}):
\begin{equation}\label{e-ph1}
{\hat H}_{\rm e-ph}\!\!=\!\!\frac{{\hbar}^{1/2}}{L L_z^{1/2}}
  \!\!\!\sum_{{\bf q},{k}_z,s}\!\!
  {U'}_{s}({\bf k})
  {\hat {\cal P}}_{{\bf k},s}
{\cal H}_{{\rm e-ph}}({\bf q})\!+
  \mbox{H.~c.},\vspace{-1.mm}
\end{equation}
where $L^2\!=\!2\pi {\cal N}_\phi l_B^2$ is the 2D area, and $L_z$ the dimension of the sample along ${\hat z}$,
\begin{equation}\label{e-ph2}
{\cal H}_{\rm e\!-\!ph}({\bf q})\!=\!\!
   \int\! e^{i{\bf qr}}
   {\hat \Psi}^{\dag}({\bf r}){\hat \Psi}({\bf r})\,d^{2}r,\quad {\bf r}=(x,y);
\end{equation}
${\hat {\cal P}}_{{\bf k},s}$ is the phonon annihilation operator (index $s$ denotes
possible phonon polarizations: longitudinal or one of two transverse polarizations), and
${U'}_{s}({\bf k})\!=\!U_{s}({\bf k})\Phi(k_z)$ is the renormalized
the vertex which includes the fields of deformation and piezoelectric
couplings. Averaging over coordinate $z$ has already being
performed and is reduced to the appearance of formfactor  \vspace{-1.mm}
\begin{equation}\label{U-renorm}
\Phi(k_z)=\int \chi^{*}(z)e^{ik_{z}z}\chi(z)\,dz.\vspace{-1.mm}
\end{equation}
The isotropic model for the phonon field$\,$\cite{gale87} enables us to take into account
deformation and piezoelectric couplings independently and use the approximation where
we take no difference between longitudinal and transverse sound velocities. In this case for the three-dimensional vertex one needs only an expression for the sum of
squares equal to $\,$
\begin{equation}\label{vertex}
 \sum_s|U_s|^2=\pi\varepsilon_{\rm ph}({\bf k})/{\cal K}_0^3\tau_{\rm A}({\bf k}),\vspace{-3.mm}
\end{equation}
where the phonon energy is $\epsilon_{\rm ph}=\hbar c_s\sqrt{k_z^2\!+\!q^2}/l_B$ (here and further both, $k_z$ and ${\bf q}$, are considered to be dimensionless), ${\cal K}_0=2.52\cdot
10^6\,$cm${}^{-1}$ is the material parameter of GaAs (see Ref.
\onlinecite{gale87}), and $\tau_{\rm A}({\bf k})$ is the characteristic time of 3D acoustic phonons, namely:
\begin{equation}\label{tau_A}
\frac{1}{\tau_{A}({\bf k})}=
  \frac{1}{\tau_D} +\frac{5({\cal K}_0l_B)^2}
  {k^6\tau_P}(
  q_x^2q_y^2+q^2k_z^2),
\end{equation}
calculated under the condition that ${\hat x},\;{\hat y},\;{\hat z}$ are the
directions of the principal crystal axes of the cubic lattice.\cite{di96,di00}
Nominal times $\tau_D\simeq 0.8\,$ps and $\tau_P\simeq 35\,$ps characterize respectively deformation and polarization electron--acoustic-phonon
scattering in the three-dimensional GaAs crystal for longitudinal sound velocity $c_s\!=\!5.14\cdot 10^5\,$sm/c
(see Ref. \onlinecite{di96} and cf. Ref. \onlinecite{gale87}).

The dimensionless operator \eqref{e-ph2} can be presented in terms of the excitonic representation in the usual way.\cite{di96,di00} In our case the relevant terms are:
\begin{equation}\label{e-ph}
{\cal H}_{{\rm e\!-\!ph},{\bf q}}={\cal N}_\phi e^{-q^2\!/4}\!\left[{\cal A}_{{\bf q}}\! + (1\!-\!q^2\!/2)
  {\cal B}_{{\bf q}}\right],
\end{equation}
where we use intra-sublevel operators
$$
\displaystyle{{\cal A}_{{\bf q}}^{\dag}\!={{\cal N}_{\phi}}^{-1}\!\sum_{p}\,
  e^{-iq_x\!p}\,
  a_{p+\frac{q_y}{2}}^{\dag}\,a_{p-\frac{q_y}{2}}}\vspace{-3.mm}
$$
(${\cal A}_{{\bf q}}\!\!\equiv\!{\cal A}_{-{\bf q}}^{\dag},\;$ ${\cal B}_{\bf q}^{\dag}$ means the $a\to b$ substitution; we note that ${\cal B}_{\bf q}^{\dag}|0\rangle\!\equiv\!0$ and ${\cal A}_{\bf q}^{\dag}|0\rangle\!\equiv\!\delta_{{\bf q},0}|0\rangle$). When so doing, comparing to  previous works,\cite{di96,di00} one has taken into account that the $a_p$ and $b_p$  operators belong to different Landau levels. Therefore, the factors in the square brackets at ${\cal A}_{{\bf q}}$ and ${\cal B}_{{\bf q}}$ in Eq. \eqref{e-ph} turn out to be Laguerre polynomials $L_0(q^2\!/2)\!\equiv\!1$ and $L_1(q^2\!/2)\!\equiv\!(1\!-\!q^2\!/2)$.

Our next task is to calculate the transition matrix element of operator \eqref{e-ph1} between the initial state $|N\rangle$ and the final one ${\hat {\cal P}}_{{\bf k},s}^\dag|N;{\bf q}_1,{\bf q}_2\rangle$,\vspace{-2.mm}
\begin{equation}\label{matrix_e_ph}
{}\!{\cal M}_{{\bf k},s,{\bf q}_1,{\bf q}_2}\!=\!\displaystyle{\frac{\langle {\bf q}_2,{\bf q}_1;\!N|{\hat {\cal P}}_{{\bf k},s}{\hat H}_{\rm e-ph}|N\rangle}{\left(\langle N|N\rangle\langle {\bf q}_2,{\bf q}_1;N|N;{\bf q}_1,{\bf q}_2\rangle\right)^{1/2}}},\vspace{-1.mm}
\end{equation}
and, thus, finding the probability of transition per unit of time according to the well-known formula\vspace{-2.mm}
\begin{equation}\label{rate}
W_{{\bf k},s,{\bf q}_1,{\bf q}_2}\!\!=\!\frac{2\pi}{\hbar}\!\left|{\cal M}_{{\bf k},s,{\bf q}_1,{\bf q}_2}\right|^2\!\delta[{\cal E}_{q_1}\!+{\cal E}_{q_2}\!+\epsilon_{\rm ph}({\bf k})]\vspace{-1.mm}
\end{equation}
($\delta[...]$ is the Dirac delta-function).
If we perform the summation
\begin{equation}\label{rate1}
R_{\bf p}=\displaystyle{\sum_{{\bf k},s,{\bf q}_2\vspace{-1.mm}}\!\!W_{{\bf k},s,{\bf p},{\bf q}_2}},\vspace{-2.mm}
\end{equation}
we obtain the total probability of transition to a state, where one of `nonzero' magnetoexcitons has a fixed wave vector: ${\bf q}_1\!=\!{\bf p}$ (or ${\bf q}_2\!=\!{\bf p}$).

Calculation of the ${\cal M}_{{\bf k},s,{\bf q}_1,{\bf q}_2}$ value is reduced to calculating the expectation\vspace{-1.mm}
\begin{equation}\label{matrix1}
M_{{\bf q},{\bf q}_1,{\bf q}_2}\!\!\!=\!\langle 0|{\cal Q}_{{\bf q}_1}{\cal Q}_{{\bf q}_2}({\cal Q}_{\bf 0})\!{}^{N\!-\!2}{\cal H}_{{\rm e\!-\!ph},{\bf q}}^\dag({\cal Q}_{\bf 0}^\dag)\!{}^N|0\rangle,\vspace{-1.mm}
\end{equation}
which is based on the commutation algebra for the ${\cal A}_{...}$-, ${\cal B}_{...}$- and ${\cal Q}_...$-operators many times used in previous works.\cite{di05,di19,di96,di00} The matrix elements, where bra- and ket- vectors are multi-exciton states, have also been numerously calculated (see, e.g., Ref. \onlinecite{di19}). In this letter we omit the details of relevant algebraic manipulations and present only the main key points of this procedure. First, note that the sum ${\cal A}_{\bf q}^\dag\!+\!{\cal B}_{\bf q}^\dag$ commutes with ${\cal Q}_{\bf 0}^\dag$, hence, the only term of operator \eqref{e-ph} contributing to expectation \eqref{matrix1} is $-{\cal N}_\phi (q^2\!/2)e^{-q^2\!/4}{\cal B}_{\bf q}^\dag$. By calculating \vspace{-1.5mm}
$$
  \displaystyle{{\cal B}_{\bf q}^\dag({\cal Q}_{\bf 0}^\dag)^N|0\rangle\equiv({N}/{\cal N}_\phi){\cal Q}_{\bf q}^\dag({\cal Q}_{\bf 0}^\dag)^{N\!-\!1}|0\rangle}\vspace{-1.5mm}
$$
and after that calculating the action of ${\cal Q}_{\bf q}$ onto state \eqref{fin}, we obtain:\cite{foot3}
\vspace{-1.mm}
\begin{equation}\label{matrix2}{}\!{}\!{}\!{}\!{}\!{}\!{}\!
\begin{array}{l}
\:M_{{\bf q},{\bf q}_1,{\bf q}_2}{}\!{}\!{}\vspace{1.mm}\\
\:=\displaystyle{\frac{N}{{\cal N}_\phi}}\,q^2e^{-q^2\!/4}\!\cos{\![{\bf q}_1\!\!\times\!{\bf q}_2]}{\vphantom{\left(\!\frac{N}{{\cal N}_\phi}\!\right)}}\langle N\!\!-\!1|N\!\!-\!\!1\rangle\delta_{{\bf q}_1\!+\!{\bf q}_2,{\bf q}}.\vspace{-1.mm}
\end{array}\vspace{-.5mm}
\end{equation}
Then we find matrix element \eqref{matrix_e_ph} and finally with the help of Eqs. \eqref{norms}, \eqref{e-ph1} and \eqref{U-renorm} -- \eqref{tau_A} perform summation in accordance with formula \eqref{rate1}. To implement this, we model formfactor \eqref{U-renorm} by employing as $\chi(z)$ a function numerically calculated with the help of a conventional routine procedure (first appeared in Ref. \onlinecite{pi92}; see also, for instance, Ref. \onlinecite{di19}), and in our case already used for calculation of the dispersion curve shown in Fig. 1. The specific function $\chi$ employed in the present work is relevant to a quantum well where the filling factor $\nu\!=\!2$ would correspond to magnetic field $B\!=\!4.18\,$T (see Ref. [\onlinecite{zh19}]). The result, $\Phi(k_z)$, is demonstrated in the inset in Fig. 2. We also model the excitation spectrum \eqref{dispersion} with polynomial:\vspace{-2.mm}
\begin{equation}\label{energy2}
{\cal E}_{q}=(e^2\!/\!\kappa l_B\!){\cal E}_m\!\left(\!2q^3\!/q_m^3-3q^2\!/q_m^2\right),\vspace{-1.mm}
\end{equation}
where ${\cal E}_m$ and $q_m$ are measured in $e^2\!/\!\kappa l_B$ and $1/l_B$ units, respectively [see Fig. 1; the graph of function \eqref{energy2} is shown by the green line; the $B$-dependencies of ${\cal E}_m$ and $q_m$, found under the condition of fixed size-quantized function $\chi(z)$ and filling factor $\nu\!=\!2$, are shown in the inset]. So,  we obtain \vspace{-1.mm}
\begin{equation}\label{RpI}
R_{p}=I(p)(N/{\cal N}_\phi)^2\!/({\cal K}_0l_B)^3,\vspace{-2.mm}
\end{equation}
where \vspace{-2.mm}
\begin{equation}\label{Rp}
{}\!{}\!{}\!{}\!\begin{array}{l}
\displaystyle{{}I(p)\!=\!\int\!\!\int dqd\varphi\frac{q^5e^{-q^2\!/2\!}}{2\pi\tau_A(q,k_z)}}\qquad\quad\vspace{1.mm}\\
\qquad\quad\times\displaystyle{|\Phi(k_z)|^2\!
\left(\frac{q^2}{k_z}\!+\!k_z\right)\cos^2{\!(pq\sin{\varphi})}}
\end{array}\vspace{-1.mm}
\end{equation}
(certainly, $R_p\!\equiv\!R_{\bf p})$. Here $k_z$ is determined by equations \eqref{momenta_save} and \eqref{energy_save}, \vspace{-1mm}
\begin{equation}\label{k_z}
{}\!{}\!k_z({\bf p},{\bf q})\!=\!\sqrt{(l_B/\hbar c_s)^2\left({\cal E}_p+{\cal E}_{|{\bf q}\!-\!{\bf p}|}\right)^2\!\!-\!q^2}\,,
\vspace{-1.mm}
\end{equation}
and also $\tau_A$ is presented by formula \eqref{tau_A} where averaging $\overline{q_x^2q_y^2}=q^4\!/8$ is performed. The area of the integration in Eq. \eqref{Rp} is determined by two conditions: (i) the energy in the final two-exciton state with nonzero momenta \eqref{fin} must be smaller than that in the initial state, and (ii)
the root expression in formula \eqref{k_z} for $k_z$ must be positive. Both can be presented by the inequality:\vspace{-1.5mm}
\begin{equation}\label{condition}
(l_B/\hbar c_s)\left({\cal E}_{p}\!\!+\!{\cal E}_{|{\bf q}\!-\!{\bf p}|}\right)+q<0.\vspace{-1.mm}
\end{equation}

The physical meaning of the value $R_p$ \eqref{RpI} is that it represents the rate of appearance of a magnetoexciton with momentum ${\bf p}$ due to the considered process of direct transition from the initial coherent state $|N\rangle$ \eqref{ini} to any state $|N;{\bf p},{\bf q}\rangle$ with unfixed number ${\bf q}$. When studying the problem kinetically and neglecting any inter-magnetoexcitonic coupling, $R_p$ will mean the rate of filling of a `one-particle' magnetoexcitonic state with specific momentum ${\bf p}$. It is obvious that the total rate induced by phonon-emission, $R\!=\!\sum_{\bf p}R_p$, is, on the one hand, the rate of the coherent state $|N\rangle$ decay${}\!$/stochastization, and, on the other hand, the rate of appearance of nonzero magnetoexcitons in the system. The physical meaning of the value R allows us to consider the kinetic equation:\vspace{-2.mm}
\begin{equation}\label{dN/dt}
dN\!/dt=\displaystyle{-R\equiv - \frac{N^2}{{\cal N}_\phi{\cal T}}},\vspace{-2.mm}
\end{equation}
where\vspace{-3.mm}
\begin{equation}\label{tau}
\frac{1}{\cal T}=\frac{1}{({\cal K}_0l_B)^3}\!\!\int_0^\infty\!\!I(p)pdp.\vspace{-2.mm}
\end{equation}
is calculated with the help of Eqs. \eqref{Rp} and \eqref{condition} (see the result in Fig. 2; the employed formfactor $\Phi(k_z)$ is demonstrated in the inset). For our specific case, when $B\!=4.18\!$T (c.f. Ref. \onlinecite{zh19}), we get numerical value ${\cal T}\!\approx\!0.88\,$ns.

Solving equation \eqref{dN/dt}, we obtain the time law of change of the number of $q\!=\!0$ excitations: \vspace{-2.mm}
\begin{equation}\label{time_law}
n(t)=n(0)\!\mbox{\Large /}\!\left[1+t\,n(0)/\!{\cal T}\right].\vspace{-1.mm}
\end{equation}
Here $n(t)\!=\!N\!/{\cal N}_\phi$ is the concentration of zero-momentum CSFEs, while value \vspace{-3.mm}
\begin{equation}\label{relative_concentration}
  1-n(t)\!/n(0)\equiv n(t)t\!/\!{\cal T}\vspace{-1.5mm}
\end{equation}
is the relative concentration of nonzero magnetoexcitons  with respect to the given total CSFE concentration $n(0)\!=\!N(0)/{\cal N}_\phi$ in the system. When dividing the `partial' rate $R_p$ by the total one $R$ we obviously obtain a `one-particle' nonzero magnetoexciton distribution function, \vspace{-2.mm}
\begin{equation}\label{f_p}
f_p=R_p/R,\vspace{-2.mm}
\end{equation}
normalized by unit, $\left(\sum_{\bf p}f_p\!=\!1\right)$. As both values, $R_p$ and $R$, have the same time dependance, $\propto\!{N(t)}^2$, the ratio $f_p$ turns out to be time-independent. However, normalization by the relative concentration of nonzero magnetoexcitons \eqref{relative_concentration} seems to be more natural. Indeed, the change from $f_p$ to $f'_p(t)\!=\! n(t)tf_p/{\cal T}$ reveals the physical meaning of the value \eqref{f_p}, namely: this one actually represents the final distribution function $f'_p(\infty)$ at $t\!\to\!\infty$, when the stochastization is completed and only non-coherent excitations with nonzero momenta are present in the system.

We neglect any correlation among appearing nonzero magnetoexcitons and their coupling with the zero ones. This concerns also any inter-excitonic correlations with participation of nonzero magnetoexcitons, including correlation induced by coupling with the lattice and so resulting in energy release (phonon emission). Thus, it has been assumed that the coherent ensemble of zero-momentum excitons \eqref{ini} is the only generator of nonzero magnetoexcitons with neglecting any subsequent evolution of the emerging nonzero magnetoexcitonic ensemble. In principle, this approach should be suitable if the relative nonzero-magnetoexciton concentration \eqref{relative_concentration} is small ($\ll\!1)$ and, in addition, the temperature is sufficiently low to ignore any phonon-absorption processes. In this case thermalization in the studied electron system should be a much longer process than the stochastization considered.

However, it is still interesting to compare the distribution function \eqref{f_p} established due to stochastization to a thermodynamically equilibrium distribution corresponding to some temperature. The latter should be Boltzmann due to the rarefaction of magnetoexiton gas [$N(0)\!\ll\!{\cal N}_\phi]$, namely:\vspace{-1.mm}
\begin{equation}\label{FT}
F_p^{(T)}=\displaystyle{e^{-{\cal E}_{p}\!/T}\!\!\!\mbox{ \huge{/}}\!\!\!\int_0^\infty\!\!\!e^{-{\cal E}_{p}\!/T}pdp}\,.\vspace{-1.mm}
\end{equation}
In Fig. 2 we demonstrate both distributions equally normalized, that is: $F_p^{(T)}$ and $F_p={\cal N}_\phi f_p$. In the $T\to 0$ limit $F_p^{(T)}\!\!\propto \delta(p-q_m)$. However, already at temperature $T\sim 0.5\,$K, the stochastization distribution $F_p$ becomes qualitatively similar to the thermodynamically equilibrium one $F_p^{(T)}$.
\vspace{-2.mm}
\begin{figure}[h]
\begin{center}
\vspace{-0.mm}
\hspace{-0.mm}
\includegraphics*[angle=0,width=.48\textwidth]{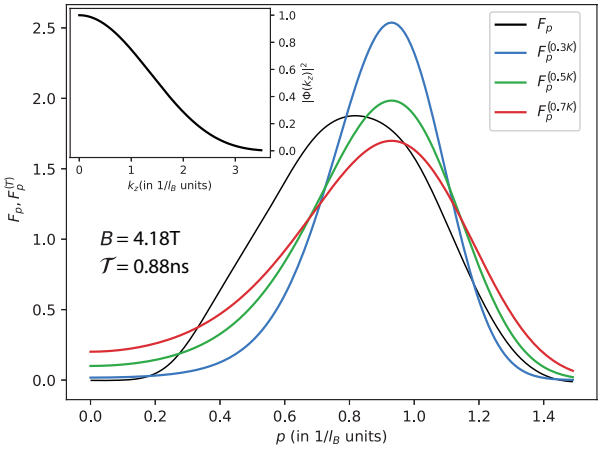}
\end{center}
\vspace{-6.mm}
\caption{The result of calculating the distribution function $F_p$ of CSFEs emerging due to the stochastization process [the black line; see Eq. \eqref{f_p} and the text], and the thermodynamically equilibrium distribution functions $F_p^{(T)}$ at different temperatures \eqref{FT}. All graphs correspond to $B=4.18\,$T. In the upper left: formfactor \eqref{U-renorm} squared as function of the dimensionless $k_z$ value is shown.}
\end{figure}\vspace{2.mm}

So, the presented model results in a nonexponential decay of the initial coherent CSFE ensemble \eqref{ini}. The time dependence of the decay \eqref{time_law} is parameterized by time ${\cal T}$ (Fig.  2.). The number of zero magnetoexcitons decreases by half during time ${\cal T}\!/n(0)$ inversely proportional to the initial CSFE concentration. A tenfold decrease will take time  $\approx\!10{\cal T}\!/n(0)$, therefore, for $n(0)\!\leqslant\!0.01$ it occurs during $\gtrsim\!1\,\mu$s [cf. the 3D characteristic electron-phonon scattering time $\tau_A$ \eqref{tau_A} which is $\sim \!0.1\,$ps if $q\!\sim\!k_z\!\sim 1$, and the CSFE lifetime which is $>\!50\,\mu$s at $B\!=\!4.18\,$T if estimated on the basis of the recent experimental data$\,$\cite{zh19}).

Generally, a single magnetoexciton is able also to drift from the ${\bf q}=0$ state  to some state with $q\!>\!0$. This weak stochastization channel exists only due to violation of the translational symmetry of the system. It is known that in modern wide-thickness quantum wells only the `smooth random potential' (SRP) represents a real reason for such violation. Actual parameters characterizing the SRP are: amplitude $\Delta\!\simeq 5-7\,$K, and correlation length $\Lambda\!\simeq 50-70\,$nm ($\gg\!l_B$). The elementary process effectively resulting in stochastization and energy release, is conversion of a zero-momentum magnetoexciton to a nonzero one with dispersion energy ${\cal E}_q\simeq -(e^2\!/\!\kappa l_B){\cal E}_m$ and an acoustic phonon with energy $\epsilon_{\rm ph}({\bf q}',k_z)=-{\cal E}_q$, where it is obvious that $q'$ is substantially less than $q$. Such a transition is described within the framework of the second order perturbation-theory approach, which represents the first order by electron-phonon coupling and simultaneously the first one by the SRP. The estimation of the characteristic time for this process gives a result that is at least $\gtrsim 10^3$  larger than the parameter ${\cal T}$ value. Thus, the single-magnetoexciton stochastization definitely ceases to be dominant even at a CSFE concentration $n(0)\!\simeq\!0.1\%$.

We note that our approach does not allow us to consider asymptotically large or small magnetic fields even if we ignore some difficulties of keeping $\nu$ equal to 2. Indeed, we are limited, first, by the condition of the two-dimensionality of the problem, i.e. the cyclotron energy should be less than the distance between the size-quantization energy levels: $\omega_c\!<\!\hbar/m^*_ed^2$, where $d^{-2}\!\!=\!\!\int\!dz|\chi(z)|^4$; and second, by the need to comply with the $r_s\!<\!1$ condition. In fact, the situation is even more complicated, because by changing the magnetic field while keeping the value of $\nu$, we must be aware that such a change corresponds to a change in the quantum heterostructure. In other words, different values of $B$ are associated with different heterostructures, therefore function $\chi(z)$ must be appropriately recalculated depending on $B$. Accordingly, the formfactors $F$ and $\Phi$ and spectrum ${\cal E}_q$ must be recalculated, too. However, in the present theoretical work, we of course, have a formal right, by keeping $\chi(z)$ and $\nu$, to calculate, at different $B$, the dispersion determined by Eqs. \eqref{dispersion} and \eqref{formfactor}. Qualitatively, the form of the dispersion curve does not change, but the ${\cal E}_q$ spectrum presented in dimensionless units becomes smoother with increasing $B$; for instance, the change of  parameters ${\cal E}_m$ and $q_m$ is shown in the inset in Figure 1. The dimensionless $k_z$ \eqref{k_z} and the integration domain \eqref{condition} depend on the field indirectly through the dependencies ${\cal E}_m(B)$ and $q_m(B)$, meanwhile, calculations show that integrals in Eqs. \eqref{Rp} and \eqref{tau} turn out to be extremely sensitive to a change in $B$ (actually to a change in the integration domain). For instance, if $B$ grows from $4$ to $8\,$T the integral in Eq. \eqref{tau} decreases by about 20 times. We emphasize that this is a formal and physically rather meaningless result, yet, this is a significant sign of the sharp dependence of our calculation on the CSFE spectrum.

At the same time, there is a hypothetical situation, although unrealistic but quite self-consistent, where there are no problems associated with the dependence of  calculation on the magnetic field.  Namely, this is an ultra two-dimensional case: when $d\!\ll\!l_B$, but the $r_s$ smallness is not violated and the $\nu\!=\!2$ condition holds. In this limit, both formfactors $\Phi(k_z)$ and $F(q)$ are set equal to unit. Then the dimensionless value $k_z$ \eqref{k_z} and the integration domain \eqref{condition} are $B$-independent. As a result, with the help of Eqs. \eqref{RpI}-\eqref{tau} [in Eqs. \eqref{k_z} and \eqref{condition} we use just analytical expression \eqref{dispersion} with $F\!=\!1$] for this ideally 2D case we find:\vspace{-1.9mm}
\begin{equation}\label{ultra2D}
1/{\cal T}\approx\left(49.4B^{3/2}+3.9B^{1/2}\right)\,\mbox{ns${}^{-1}$},\vspace{-1.9mm}
\end{equation}
where $B$ is in Tesla. The first term here comes from the deformation electron-phonon interaction and the second one from the {\em e-ph} polarization coupling [see. Eq. \eqref{tau_A}]. This formula for $B\!=\!4.18\,$T gives value ${\cal T}$ about 400 times smaller than our calculation above performed for a real physical system.

In conclusion, we note another feature of the stochastization channel under consideration. We have seen already that integral \eqref{Rp} and therefore the obtained result \eqref{tau} is very sensitive to changes in the condition \eqref{condition} for the phase volume of the occurring stochastization processes and, hence, to any change of the energy dispersion spectrum (Fig. 1). However, we do not have experimental data on this issue, and theoretical considerations leading to the employed dependence [Eqs. \eqref{dispersion}, \eqref{energy2} and Fig. 1] contain significant simplifications. In particular, we neglected any $q$-dependence of the negative Coulomb shift associated with the second-order Coulomb correction $\sim\! r_s^2\hbar\omega_c$ to the energy spectrum (see above). We ignored also any inter-magnetoexcitonic corrections. In the coherent state, these virial corrections to the CSFE spectrum can be negative and more significant than in the incoherent ensemble.\cite{di19} Then, effectively, when recalculated by one excitation, the dependence ${\cal E}_q$ will be more smoothed. Both mentioned corrections can reduce the $({\bf p},{\bf q})$-domain of the integration in Eqs. \eqref{Rp} and \eqref{tau}, and thus lead to increasing the stochastisation time.

The authors are grateful to L.V. Kulik for useful discussions. The research was supported by the Russian Science Foundation: grant RSF-21-12-00386.


\vspace{-5.mm}
\begin{thebibliography}{99}

\bibitem{UFN}
L.V. Kulik, A.V. Gorbunov, S. Dickmann, V.B. Timofeev, Phys. Usp. {\bf 62}, 865 (2019).

\bibitem{ka84}
{C. Kallin and B.I. Halperin}, Phys. Rev. B {\bf 30}, 5655 (1984).

\bibitem{di05}
S. Dickmann and I.V. Kukushkin, Phys. Rev. B {\bf 71}, 241310(R) (2005).

\bibitem{ku05}
L.V. Kulik, I.V. Kukushkin, S. Dickmann, V.E. Kirpichev, A.B. Van'kov, A.L. Parakhonsky, J.H. Smet, K. von Klitzing, W. Wegscheider, Phys. Rev. B {\bf 72}, 073304 (2005).

\bibitem{foot1}
This statement concerns also special phonon modes (optical or acoustic) arising in the GaAs/AlGaAs interface.

\bibitem{di19} S. Dickmann, L.V. Kulik, V.A. Kuznetsov,  Phys. Rev. B {\bf 100}, 155304 (2019).

\bibitem{foot2}
For the first time this operator was used in work:
{A.B. Dzyubenko and Yu.E. Lozovik}, Sov. Phys. Solid State {\bf 25}, 874 (1983).

\bibitem{an82}
T. Ando, A.B. Fowler, and F.Stern, Rev. Mod. Phys. {\bf 54},
437 (1982).

\bibitem{di96}S. Dickmann and S.V. Iordanskii, JETP {\bf 83}, 128 (1996).

\bibitem{di00}
S. Dickmann, Phys. Rev. B {\bf 61}, 5461 (2000).

\bibitem{di12}
{S. Dickmann and T. Ziman}, Phys. Rev. B {\bf 85}, 045318 (2012).

\bibitem{io89} S.V. Iordanskii and B.A. Muzykantskii, JETP {\bf 69}, 1006 (1989).

\bibitem{gale87}
{V. F. Gantmakher, and Y. B. Levinson}, {\em Carrier {S}cattaring in
{M}etals and {S}emiconductors} (North-Holland, Amsterdam, 1987).

\bibitem{foot3}
{In this calculation one used the fact that the expectation
$\left\langle 0|({\cal Q}_{\bf 0})^{N\!+\!1}({\cal Q}_{\bf 0}^\dag)^{N\!-\!1}{\cal Q}_{-{\bf q}_1}^\dag{\cal Q}_{{\bf q}_1}^\dag|0\right\rangle$
at ${\bf q}_1\!\neq\!0$ is independent of ${\bf q}_1$.}

\bibitem{pi92}
A. Pinczuk, B.S. Dennis, D. Heiman, C. Kallin, L. Brey, C. Tejedor, S. Schmitt-Rink, L.N. Pfeiffer, and K.W. West, Phys. Rev. Lett. {\bf 68}, 3623 (1992).

\bibitem{zh19}
A.S. Zhuravlev, V.A. Kuznetsov, A.B. Gorbunov, L.V. Kulik, V.B. Timofeev, and I.V. Kukushkin, JETP Lett. {\bf 110}, 284 (2019).

\end{thebibliography}
\end{document}